\documentclass[groupedaddress,twocolumn,preprintnumbers,amsmath,amssymb]{revtex4}
\usepackage{graphicx}
\usepackage{dcolumn}
\usepackage{bm}
\usepackage{braket}
\usepackage{color}

\begin{document}

\title{Dissipation-assisted coherence formation in a spinor quantum gas}


\author{Yujiro Eto$^{1,2}$}
\author{Hitoshi Shibayama$^{3}$}
\author{Kosuke Shibata$^{4}$}
\author{Aki Torii$^{4}$}
\author{Keita Nabeta$^{4}$}
\author{Hiroki Saito$^{5}$}
\author{Takuya Hirano$^{4}$}
\affiliation{%
$^{1}$National Institute of Advanced Industrial Science and Technology (AIST), NMIJ, Tsukuba, Ibaraki 305-8568, Japan\\
$^{2}$JST, PRESTO, 4-1-8 Honcho, Kawaguchi, Saitama 332-0012, Japan\\
$^{3}$College of Industrial Technology, Nihon University, Narashino 275-8576, Japan\\
$^{4}$Department of Physics, Gakushuin University, Toshima, Tokyo 171-8588, Japan\\
$^{5}$Department of Engineering Science, University of Electro-Communications, Chofu, Tokyo 182-8585, Japan}

\date{\today}
             
\maketitle
{\bf 
Dissipation affects all real-world physical systems and often induces energy or particle loss, limiting the efficiency of processes.
Dissipation can also lead to the formation of dissipative structures or induce quantum decoherence.
Quantum decoherence and dissipation are critical for quantum information processing. 
On the one hand, such effects can make achieving quantum computation much harder, but on the other hand, dissipation can promote quantum coherence and offer control over the system. 
It is the latter avenue –- how dissipation can be exploited to promote coherence in a quantum system –- that is explored in this work.
We report the exploration of dissipation in a Bose-Einstein condensate (BEC) of spin-2 $^{87}$Rb atoms. 
Through experiments and numerical simulations, we show that spin-dependent particle dissipation can give rise to quantum coherence and lead to the spontaneous formation of a magnetic eigenstate.
Although the interactions between the atomic spins are not ferromagnetic, the spin-dependent dissipation enhances the synchronization of the relative phases among five magnetic sublevels, and this effects promotes magnetization.
}

Dissipation is a ubiquitous phenomenon in the real world: Moving objects exposed to friction ultimately stop dissipating kinetic energy into the surrounding environment. 
With regard to the present experiment using cold atoms, loss of atoms from the trap is inevitable.
However, dissipation induces not only energy or particle loss but also various interesting effects, such as quantum decoherence \cite{Streltsov17} and the formation of a reproducible steady state, called dissipative structures \cite{Glansdorff}, by the exchange of energy and particles with the environment. 

A deeper understanding of the role of dissipation leads to a deeper understanding of physical system. 
For example, the loss of quantum coherence of a superposition state of a quantum system lies at the heart of the fundamental question of how a classical world, in which a coherent superposition of macroscopic states is never observed, can be derived from quantum mechanics \cite{Streltsov17}. 
The simplest model that can be used to study this question is a quantum two-state model in which the two states are coupled to an infinite set of quantum harmonic oscillators  \cite{Leggett87}. 
Such coupling generates fluctuations in the system and leads to dissipation and decoherence; in this model dissipation and decoherence are associated. 

The question of quantum decoherence is also of great significance for practical applications such as quantum information processing  \cite{Zurek03}. 
From a computational point of view, although energy dissipation is fundamentally required for computation to discard information, dissipation and decoherence are major obstacles to realizing quantum computation because they disrupt quantum mechanical interference between different computational trajectories  \cite{Landauer}. 
On the other hand, dissipation sometimes has the completely opposite effect of promoting quantum coherence; thus, such dissipation can be used as a new control strategy for quantum systems and as a useful resource for quantum computations \cite{Verstraete09}. 
Through proper design of the coupling between the system and the surrounding environment, it is possible to prepare a desirable pure quantum state in an open quantum system to explore quantum simulation using strongly correlated many body states \cite{Diehl08}. 
Experimentally, the stabilization of entangled qubits in superconductors \cite{Shankar13, Lin13} and the control of quantum phase transitions in cold atoms \cite{Tomita17} have been demonstrated by exploiting controllable dissipation. 
However, no experiments have been reported in a quantum many-body system in which a coherent final state is generated from a qualitatively different initial state due to dissipation. 
The influence of dissipation on quantum coherence is also significant in biological systems  \cite{Lambert13}. 
In photosynthetic systems, dissipation supports quantum coherence, thereby enabling efficient electron transport subject to the dissipative environment \cite{Engel07, Ishizaki09}. 
Exploring new roles of dissipation using a highly controllable system may help to understand such an efficient process in nature.

In this paper, we investigate the dynamics of atomic spins in a dissipative quantum-degenerate gas using Bose-Einstein condensates (BECs) of spin-2 ($F = 2$) $^{87}$Rb atoms and
uncover a new role of particle dissipation in the quantum coherence.
The atomic BEC is a quantum many-body system with high controllability and we have found that the inherent dissipation plays an important role in forming coherence and generating a final state having completely different symmetry from an initial state.
Particle dissipation inevitably occurs in this system: atoms escape from the system when the transition from $F = 2$ to $F = 1$ occurs due to inelastic collisions \cite{Tojo}.
We observe the emergence of symmetry-breaking magnetization from an unpolarized spin state.
Symmetry-breaking magnetization has been observed in a spin-1 $^{87}$Rb BEC \cite{Sadler06}, in which the magnetization is induced by the ferromagnetic interactions between atoms. 
By contrast, although the corresponding interactions are not ferromagnetic in a spin-2 $^{87}$Rb BEC \cite{Klausen01}, the unpolarized spin state evolves into the ferromagnetic state due to the dissipation of atoms.
This magnetization mechanism lies in stark contrast to the conventional one, in which the spin vectors align in the same direction to lower the interaction energy \cite{Sadler06}.
Through comparisons with numerical simulations of the Gross-Pitaevskii (GP) equation, we reveal that the spin-dependent inelastic loss of atoms assists in the formation of the fully polarized ferromagnetic spin state.
Furthermore, we numerically show that not only the ferromagnetic state but also other magnetic states such as the cyclic state can be formed due to the spin-dependent dissipation of atoms.

\begin{figure}[tbp]
\includegraphics[width=9cm]{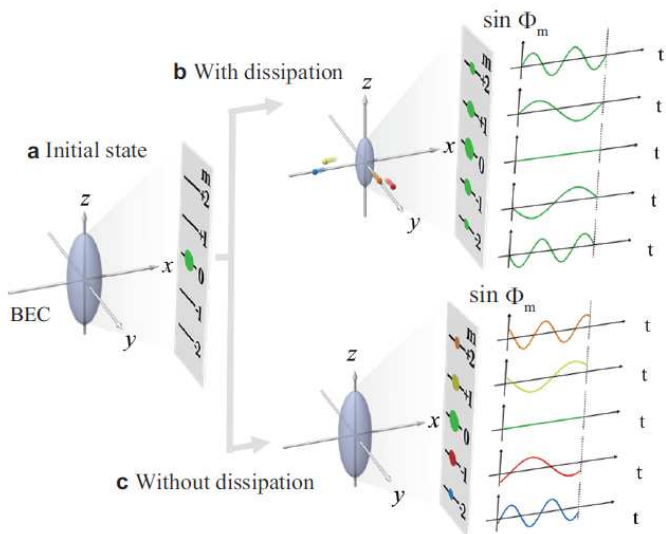}
\caption{{\bf Conceptual diagram of coherence formation assisted by spin-dependent particle dissipation.}
{\bf a,} The as-prepared unpolarized spin state of $\ket{F=2, m = 0}$,
where $F$ and $m$ are the quantum numbers for the total angular momentum and the Zeeman sublevel, respectively.
The populations in the $\ket{2,m}$ states are depicted in the grey box.
{\bf b} and {\bf c} show the spin states after time evolution with and without dissipation, respectively.
The sinusoidal curves represent the phases, $\sin{\Phi_m} = \sin{(m \omega t + \phi_m)}$, under a magnetic field in the $z$ direction.
The phase offset, $\phi_m$, is expressed by the difference in color.
Atoms populate in all five components, and their relative phases are synchronized only in the dissipative system.}
\label{f:fig1}
\end{figure}

A schematic of the observed spinor dynamics is presented in Fig. 1.
We performed the experiments using a spin-2 BEC consisting of the five magnetic sublevels, specifically, $m = -2, -1, 0, +1,$ and $+2$.
The magnetization, $S_i$, along the $i$-axis was obtained from the population in each magnetic sublevel, $\rho_{m,i}$, with the $i$-axis as the quantization axis ($S_i = \Sigma_{m} m\rho_{m,i}$).
The atoms were initially prepared in the $\ket{F = 2, m = 0}$ state with the quantization axis along a magnetic bias field on the $z$-axis.
This state is the completely unpolarized spin state with rotational symmetry around the $z$-axis (Fig. 1a).
After time evolution, the atoms were distributed into all $m$ components by two-body elastic collisions while preserving a longitudinal magnetization $S_z$ of zero.
Similar dynamics of $\rho_{m,z}$ have previously been investigated experimentally \cite{Schmaljohann04,Chang04, Kuwamoto04}, 
but relative to that in our previous experiment \cite{Kuwamoto04}, the purity of the $\ket{2,0}$ state in this study was greatly  enhanced to improve rotational symmetry.
The most significant difference from previous studies is that we also measured the transverse magnetization $S_\perp$ orthogonal to the $z$-axis, 
making it possible to obtain the information on the phase coherence.
We found the transverse magnetization to be almost fully polarized, $S_\perp \simeq 2$, after time evolution.
This result indicates that with the $z$-axis as the quantization axis, the phases in the magnetic sublevels, $\Phi_m = m \omega t + \phi_m$ rotate synchronously, i.e., the five phase offsets $\phi_m$ become the same (see the sinusoidal curves in Fig. 1b), where $\omega$ is the linear Zeeman frequency.
On the other hand, phase synchronization does not occur in a GP simulation without dissipation (Fig. 1c); thus, it can be concluded that the phase synchronization is due to dissipation.

\begin{figure}[tbp]
\includegraphics[width=7cm]{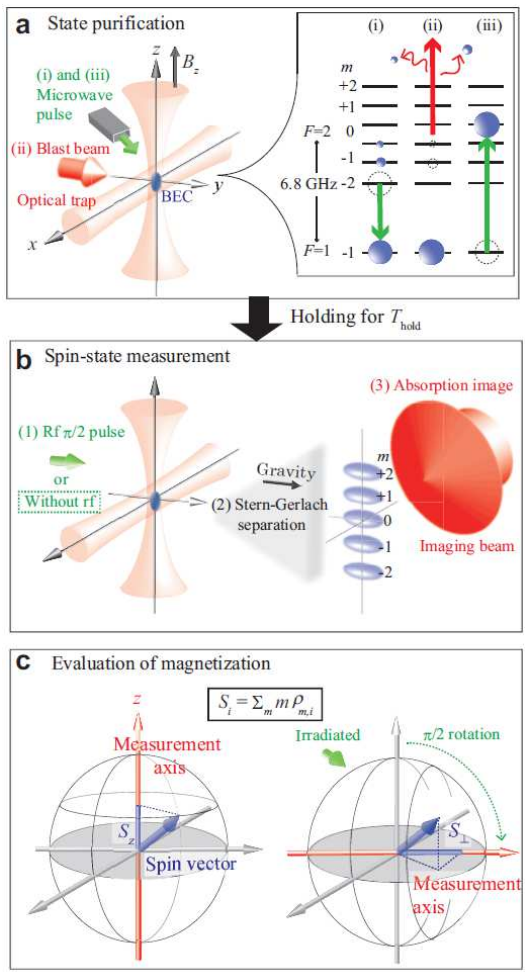}
\caption{
{\bf Experimental procedure used to prepare and measure the spin states.}
{\bf a, State purification} Generation of a pure $\ket{2,0}$ state. The BEC was irradiated with microwave and optical fields in procedures (i) to (iii), and the corresponding spin states are depicted in the right panel. 
{\bf b, Spin-state measurement} Measurement of the atomic distribution of each $m$ component.
(1) Application of a $\pi/2$ rf pulse was applied (only when measuring the spin state in the plane orthogonal to the quantization axis).
(2) Stern-Gerlach separation during the time of flight.
(3) Absorption imaging.  
{\bf c, Evaluation of magnetization} Calculation of the longitudinal and transverse magnetizations per atom $S_z$ and  $S_\perp$.
By applying (not applying) the $\pi/2$ rf pulse depicted in b(1), the value of $S_\perp$ ($S_z$) was obtained.}
\label{f:fig2}
\end{figure}

The experimental procedure used to prepare and measure the spin states is depicted in Fig. 2.
We used a BEC trapped in a crossed far-off-resonance optical trap (FORT).
The procedure shown from (i) to (iii) on the right panel in Fig. 2a was applied using microwave pulses and a blasted beam to create the pure $\ket{2,0}$ state from the state occupying the multiple $m$ levels (see Methods).
The BEC prepared in the $\ket{2,0}$ state was then held in the optical trap for a variable time of $T_{\rm hold}$.
To measure $S_z$, the BEC was released from the trap, and each $m$ component was spatially separated along the $z$ direction using the Stern-Gerlach method (Fig. 2b).
On the other hand, to measure $S_\perp$ orthogonal to the $z$ direction, we irradiated the BEC with a $\pi/2$ radio frequency (rf) pulse just before releasing it from the trap.
The $\pi/2$ rf pulse effectively rotated the measurement axis by $\pi/2$  (Fig. 2c).
After a time of flight of $15$ ms, the spatial distribution of each $m$ component was measured using absorption imaging.
The number of atoms in each $m$ component was estimated by performing the bimodal fitting of the atomic distribution.
The longitudinal or transverse magnetization per atom, $S_i$, was calculated from $\rho_{m,i}$.

\begin{figure}[tbp]
\includegraphics[width=7cm]{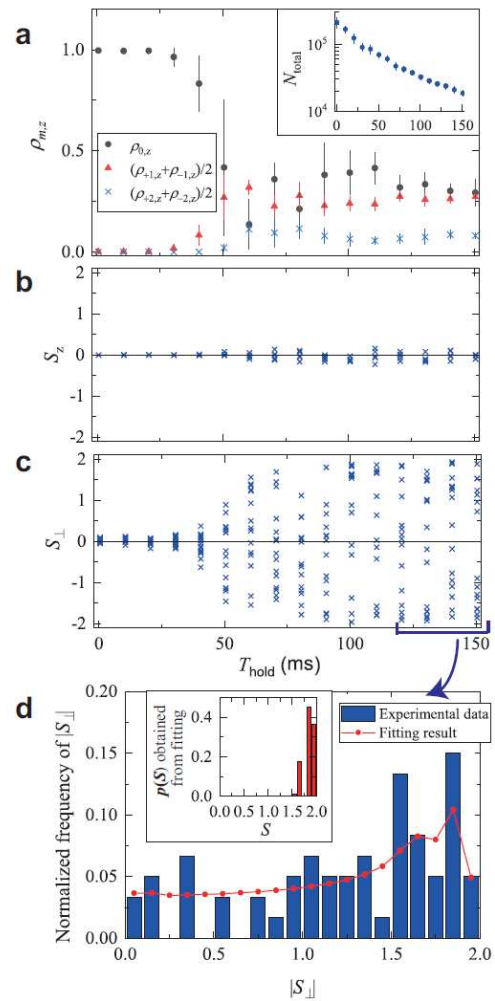}
\caption{
{\bf Observation of the spinor dynamics starting from an unpolarized spin state.}
{\bf a,} Time evolution of the populations in the magnetic sublevels $m$.
The inset shows the total number of atoms.
{\bf b,} Dynamics of the longitudinal magnetization, $S_{z}$.
{\bf c,} Dynamics of the transverse magnetization,  $S_\perp$.
{\bf d,} Normalized occurrence distribution of $\left| S_\perp \right|$ obtained during $T_{\rm hold} = 120$ - $150$ ms.
The histogram with a bin width of 0.1, was constructed from 60 experimental data points in c, and the circles indicate the fitting results (see Methods). 
The inset shows the estimated frequency distribution of $S$.
}
\label{f:fig3}
\end{figure}

We first investigate the spinor dynamics along the longitudinal axis ($z$-axis) when the $\pi/2$ rf pulse was not applied in procedure (1) in Fig. 2b.
Figure 3a shows the dependence of $\rho_{m,z}$ on the free-evolution time, $T_{\rm hold}$.
The $m = \pm1$ components grow after $T_{\rm hold} \simeq 30$ ms, followed by a delayed growth in the $m = \pm 2$ components after $T_{\rm hold} \simeq 50$ ms.
The resulting spin state consists of all five $m$ components. 
The longitudinal magnetization $S_z$ calculated from the results shown in Fig. 3a remains almost stable around zero, as shown in Fig. 3b.
As shown in the inset of Fig. 3a, the total number of atoms is reduced because this system exhibits particle dissipation due to hyperfine changing inelastic collisions \cite{Tojo}.

The initial slow rise of the $m = \pm1$ components in Fig. 3a reflects the metastability of the $m = 0$ state \cite{Schmaljohann04, Gerving12}.
Subsequent dynamical evolution in $\rho_{m,z}$ is caused by both spin-exchanging $s$-wave interaction and spin-dependent particle dissipation.
These processes change the populations $\rho_{m,z}$ while preserving the longitudinal magnetization $S_z$, since the the interaction between atoms has rotational symmetry \cite{Tojo}.

We next investigate the spinor dynamics in the transverse plane when the $\pi/2$ rf pulse is applied in the procedure (1) in Fig. 2b.
As in the case of the dynamics measured along the longitudinal axis, the initial state is almost fully maintained up to $T_{\rm hold} \simeq 30$ ms. 
However, unlike in the dynamics of $\rho_{m,z}$, the shot-to-shot variations in $\rho_{m,\perp}$ for each $T_{\rm hold}$ greatly increase after $T_{\rm hold} \simeq 30$ ms, which results in $S_\perp$ varying between $-2$ and $+2$, as shown in Fig. 3c.
The experimental results in Fig. 3c indicate that a transverse magnetization has emerged after $T_{\rm hold} \simeq 30$ ms.
The shot-to-shot variations in $S_\perp$ in Fig. 3c can be explained by a random azimuthal angle $\alpha$ of the magnetization.
Since the frequency of Larmor precession is proportional to the magnitude of the magnetic bias field, fluctuations in the magnetic field give rise to a random value for $\alpha$.
However, even if the magnetic field noise were to be completely suppressed, $\alpha$ should vary randomly due to the spontaneous breaking of the axial symmetry of the initial state.

We estimated the magnitude of the magnetization, $S= \sqrt{S_x^2+S_y^2}$, in the resulting spin state from 60 experimental data points of $\left| S_\perp \right|$ obtained during $T_{\rm hold} = 120$-$150$ ms (histogram in Fig. 3d).
We assume that $S$ varies for each measurement and that the variation obeys the probability distribution, $p(S)$ (see Methods).
The estimated $p(S)$ is plotted in the inset of Fig. 3d, where $p(S)$ is a discrete probability distribution with a bin width of $0.1$.
The expectation value and the variance of the magnetization calculated from $p(\left| S \right|)$ is $1.85\pm0.15$, which is close to the fully magnetized value $|S| = 2$.

\begin{figure}[tbp]
\includegraphics[width=8cm]{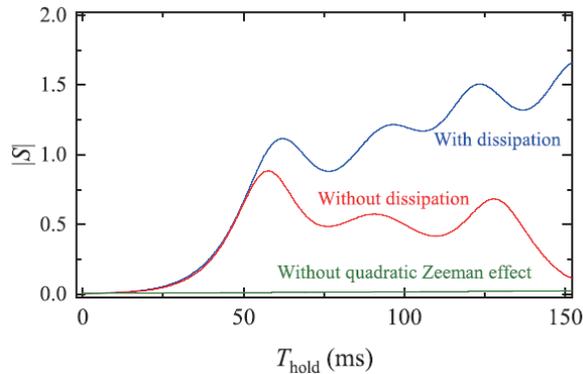}
\caption{{\bf Evolution of the magnetization from numerical simulation.} 
The blue and red curves represent simulations performed by solving the 3D GP equation with and without the dissipation of atoms, respectively.
In the green curve, the quadratic Zeeman effect is excluded. 
}
\label{f:fig4}
\end{figure}

To corroborate that the observed magnetization is caused by the particle dissipation, we performed three-dimensional numerical simulations of the GP equation (see Methods).
First we considered the case in which particle dissipation is not included in the GP equation.
Figure 4 shows a typical example of the time evolution of the transverse magnetization $S$.
As shown by the red curve, the magnetization initially grows but then decreases, with the magnitude of the magnetization being limited to no more than $S \simeq 1$, which is in contrast to the experimental findings.
This temporal growth in the magnetization is due to the spin-exchange interaction and the quadratic Zeeman effect.
The energies of the $m=\pm1$ and $\pm2$ states are lowered by the quadratic Zeeman energy $q m^2 (q < 0)$, and the transitions from the $m = 0$ state to these states are enhanced. 
The quadratic Zeeman effect also rotates the phase of each sublevel by $(q m^2 / \hbar) t$, which changes the relative phases $\phi_m - \phi_{m-1}$, resulting in a decrease in $S$ after $T_{\rm hold} \simeq 60$ ms.
When the quadratic Zeeman effect is removed, magnetization is never produced (green curve in Fig. 4).

The blue curve in Fig.~4 shows the transverse magnetization obtained by solving the GP equation with particle dissipation included.
The increase in the magnetization up to $T_{\rm hold} \simeq 60$ ms is similar to that in the case without particle dissipation.
However, the magnetization then continues to increase in the dissipative system, in agreement with the experimental result.
The continuous growth in the transverse magnetization after $T_{\rm hold} \simeq 60$ ms observed in the blue curve in Fig.~4 indicates that the dissipation synchronized the relative phases $\phi_m - \phi_{m-1}$ opposing to the phase rotation by the quadratic Zeeman effect.

The magnetization mechanism is understood from the spin-dependent part of the mean-field energy,
\begin{equation} \label{Espin}
E_{\rm spin} = \int d\bm{r} \left( \frac{g_4 - g_2}{14} \bm{s} \cdot
\bm{s} + \frac{7g_0 - 10g_2 + 3g_4}{14} |A_0|^2 \right),
\end{equation}
where $g_{\cal F}$ is the interaction coefficient for the colliding channels of
total spins ${\cal F} = 4$, 2, and 0; $\bm{s}$ is the magnetization density; and $|A_0|$ is the spin-singlet density.
The spin-dependent particle dissipation is expressed by the non-positive imaginary part of the interaction coefficient $g_{\cal F}$ in Eq.~(\ref{Espin}) \cite{Tojo}.
The mechanism driving the growth of the magnetization with the synchronization of the $\phi_m$ can be explained by the term proportional to $\bm{s} \cdot \bm{s}$ in Eq.~(\ref{Espin}).
The decay through the ${\cal F} = 4$ channel is prohibited by the conservation of angular momentum (${\rm Im}g_4 = 0$); therefore, the imaginary part of the coefficient of $\bm{s} \cdot \bm{s}$ is always non-negative, 
and the polarized state $(|\bm{s}| > 0)$ is more likely to survive than the other states.
Thus, the spin-dependent particle dissipation assists in the formation of the ferromagnetic state, which is a magnetic eigenstate.
Because the emergence of $S_z$ is prohibited by the conservation of $S_z$, the transverse magnetization emerges through this mechanism.
Thus, the dissipation synchronizes the phases $\phi_m$ on the quantization axis along the longitudinal direction giving rise to the dissipation-induced magnetization.

\begin{figure}[tbp]
\includegraphics[width=8cm]{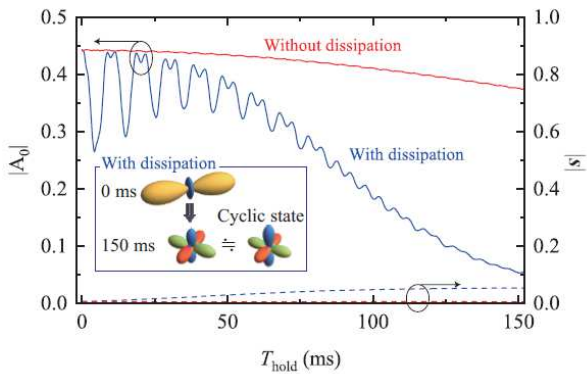}
\caption{{\bf Formation of the cyclic state (magnetic eigenstate).} 
Numerical simulation of the time evolution of $\left| A_0 \right|$ and $\left| \bm{s} \right|$, when Im$g_0$ and Im$g_2$ are ten times and one-tenth their values for $^{87}$Rb, respectively. 
The initial spin state is $\hat R_x(\pi / 2) \ket{2, 0}$. 
The size of the magnetic field, $B_z$, is $50$ mG.
The spherical harmonic representations of the spin states at the centre of the condensate are shown in the insets (see Methods).
}
\label{f:fig5}
\end{figure}

In analogy with the dissipation-induced generation of the ferromagnetic state, we expect that dissipative phase-synchronization could also assist the formation of other magnetic eigenstates if the value of the imaginary part of the interaction coefficient $g_{\cal F}$ is selected appropriately.
One such eigenstate is known as the cyclic state, which is a BEC of singlet trios of spin-2 atoms \cite{Koashi00, Ciobanu00, Klausen01, Ueda02}.
The cyclic state emerges when the singlet-pair density $|A_0|$ and the magnetization $\bm{s}$ are both suppressed. 
As mentioned above, the imaginary part of $g_4 - g_2$ in Eq. (1) is always non-negative, which causes $|\bm{s}|$ to increase. 
If the imaginary part of $7 g_0 - 10 g_2 + 3 g_4$ is large and negative, then $|A_0|$ can decrease before $\bm{s}$ grows, leading to the formation of a cyclic state. 
Figure 5 shows the time evolution of $| \bm{s} |$ and $|A_0|$ for such a case of Im$g_0$ and Im$g_2$. 
The initial state is $\hat R_x(\pi / 2) |2, 0 \rangle$, where the operator $\hat R_x(\theta)$ rotates the spin state around the $x$-axis by an angle $\theta$. 
The blue curve in Fig. 5 shows that $|A_0|$ decays to almost zero in the presence of particle dissipation while $|\bm{s}|$ remains small, indicating that the cyclic state is formed.
The spherical harmonic representations of the spin states also show that the initial polar state evolves into the cyclic state, as shown in the inset of Fig. 5.
Notably that such tetrahedral symmetry arises from the particle dissipation.
Both the initial state and the cyclic state have populations in the $m = 0$ and $m = \pm 2$ states, with relative phases of $\phi_2 + \phi_{-2} - 2 \phi_0 = 0$ for the initial state and $\pi$ for the cyclic state. 
Thus synchronization due to spin-dependent particle dissipation also occurs in the generation of cyclic state.

In conclusion, we have experimentally and theoretically investigated the role of dissipation in a BEC of spin-2 $^{87}$Rb atoms and found that particle dissipation can give rise to quantum coherence.
Although the interactions between the atomic spins are not ferromagnetic, 
we nonetheless observed the emergence of transverse magnetization, in which the relative phases among the five magnetic sublevels are synchronized. 
Numerical simulations revealed that this phenomenon is due to spin-dependent particle dissipation.
It has also been shown that with appropriate loss parameters, such dissipation can lead to the formation of a cyclic magnetic state.
These results indicate that naturally occurring dissipation gives rise to robustness of quantum coherence.
The existence of robust coherence under the highly dissipative environment is of great interests in the field of quantum biology \cite{Lambert13}.
Therefore, exploring the effects of dissipation using highly controllable system such as the atomic BEC may promote understanding such real world quantum phenomena.

\section*{Methods}
\noindent {\bf Preparation and measurement.}
The experimental configuration is the same as in our previous work \cite{Eto18}.
The BEC of the $\ket{2,2}$ state was created by means of rf evaporative cooling in a magnetic trap.
The magnetically-trapped BEC was loaded into a FORT to liberate the spin degrees of freedom.
The axial and radial frequencies of the FORT were $\omega_z / (2\pi) = 64$ Hz and $\omega_r / (2\pi) = 190$ Hz, respectively.
An external magnetic field of $B_z = 200.9$ mG was aligned with the axis of the FORT trap (the $z$-axis), to produce a quadratic Zeeman shift, $|q|/h \simeq 3$ Hz. 
The stable magnetic field environment is required to accurately control the spin state.
Therefore, the entire experimental configuration was installed inside a magnetically shielded room with walls of permalloy plates.
In addition, laser diode sources with a low ripple noise of less than 1 $\mu$A (Newport 505B) were used as the current source for the generation and compensation of the magnetic field.
Approximately 90\% of the atoms loaded into the optical trap occupied the $\ket{2,-2}$ state from the magnetically trapped BEC. 
The remaining $\simeq$ 10\% of the atoms, in the $\ket{2,-1}$ and $\ket{2,0}$ states, needed to be eliminated to create a BEC occupying the single $\ket{2,0}$ state.
For this purpose, as shown in Fig. 2a, after the $\ket{2, -2}$ atoms were stored in the $\ket{1,-1}$ state using a resonant microwave pulse, and
the atoms in the $F = 2$ manifold were blasted from the trap using a beam resonant with $F = 2$ to $F' = 3$.
A BEC in the $\ket{2,0}$ state was created by transferring atoms from the $\ket{1,-1}$ state back into the $\ket{2,0}$ state, which then contained 3 $\times 10^{5}$ atoms with a condensate fraction of $\simeq 70$ \%.
The rf $\pi/2$ pulse to switch the measurement axis was temporally shaped by the Gaussian form using an arbitrary-waveform generator (Keysight Technologies, 33522B). 
The standard deviation of Gaussian rf envelope was 11.5 $\mu$s.
After the time evolution,the BECs are released from the FORT to measure the atom number of each $m$ component.
An absorption image was acquired using a linearly polarized light resonant with $F =2$ - $F' = 3$.

\noindent {\bf Model for occurrence distribution of $\left| S_\perp \right|$.}
The probability that $\left| S_\perp \right|$ takes a value between $a$ and $a+\delta a$ can be expressed as 
\begin{eqnarray} \label{M}
\int_{0}^{2} d S p({S}) \int_{a}^{a+\delta a} d s f(s, S),
\end{eqnarray}
where $f (s, S) =  (\pi^2 (S^2-s^2))^{-1/2}$.
In this model, $S$ obeys the probability distribution $p(S)$ in each measurement.
The circles in Fig. 3d indicate the values calculated using $p(S)$ such that the square sum of the difference between the experimental values (bars in Fig. 3d)  and Eq.~(\ref{M}) is minimized.

\noindent {\bf Theoretical procedure.} 
In the mean-field approximation, the macroscopic wave functions $\psi_m(\bm{r}, t)$ of the $| F = 2, m \rangle$ state obey the GP equation given by $i\hbar \partial \psi_m / \partial t = \delta E / \delta \psi_m^*$.
The total energy $E$ takes the form,
\begin{eqnarray} \label{E}
E & = & \int d\bm{r} \sum_m \psi_m^* \left( -\frac{\hbar^2}{2M} \nabla^2
+ V_{\rm trap} + p m + q m^2 \right) \psi_m
\nonumber \\
& & + \frac{1}{2} \int d\bm{r} \biggl( \frac{4g_2 + 3g_4}{7} \rho^2
+ \frac{g_4 - g_2}{7} \bm{s} \cdot \bm{s}
\nonumber \\
& & + \frac{7g_0 - 10g_2 + 3g_4}{7} |A_0|^2 \biggr),
\end{eqnarray}
where $M$ is the mass of a $^{87}{\rm Rb}$ atom, $V_{\rm trap}$ is the trap potential, and $p$ and $q$ are the linear and quadratic Zeeman coefficients, respectively.
In Eq.~(\ref{E}), we define $\rho = \sum_m |\psi_m|^2$, $\bm{s} = \sum_{mm'} \psi_m^* \bm{s}_{mm'} \psi_{m'}$, and $A_0 = (2 \psi_2 \psi_{-2} - 2 \psi_1 \psi_{-1} + \psi_0^2) / \sqrt{5}$, where $\bm{s}$ is the vector of the spin-2 matrices.
According to Ref.~\cite{Tojo}, the inelastic collisional loss can be incorporated into the mean-field approximation through the imaginary part of the interaction coefficient as follows: $g_{\cal F} = 4 \pi \hbar^2 a_{\cal F} / M
- i \hbar b_{\cal F} / 2$, where $a_{\cal F}$ and $b_{\cal F} \geq 0$ are the $s$-wave scattering length and loss rate, respectively, for the colliding channel of total spin ${\cal F}$.
The values of $a_{\cal F}$ and $b_{\cal F}$ are taken from
Refs.~\cite{Tojo,Widera}.
Loss through the ${\cal F} = 4$ channel is prohibited; thus $b_4 = 0$.

The spherical harmonic representation of the spin state used in Fig. 5 is defined as
\begin{equation}
S(\bm{\Omega}) = \sum_{m=-F}^F \frac{\psi_m}{\sqrt{\rho}} Y_F^m(\bm{\Omega}),
\end{equation}
where $\bm{\Omega}$ is the direction in three-dimensional space and the $Y_F^m$ are the spherical harmonics.
This representation clearly visualizes the symmetry of the spin state with respect to rotation in spin space.

\section*{Acknowledgments}
This work was supported by JSPS KAKENHI (grant nos. JP17K05595, JP17K05596, JP16K05505, JP15K05233, and JP25103007) and JST PRESTO (grant no. JPMJPR17G3).
Y.E. acknowledges support from the Leading Initiative for Excellent Young Researchers (LEADER).

\section*{Author contributions}
Y. E., H. S., and T. H. conceived the research project. 
Y. E. designed the experiments.
Y. E., H. S., A. T., and K. N. acquired and analysed the data.
H. S. designed the theoretical framework and performed the theoretical calculations.
Y. E., K. S., H. S., and H. T. discussed and interpreted the experimental and theoretical results.
All of the authors contributed to writing the manuscript.

\section*{Competing interests}
The authors declare no competing interests.

\newpage

\end{document}